\documentclass[11pt,twoside]{article} 
\usepackage{asp2004}
\usepackage{epsf}
\usepackage{psfig}
\usepackage{lscape} 

\begin{document} 
\title{Non-Ideal Equation of State, Refraction and Opacities in Very Cool, Helium-Rich White Dwarf Atmospheres}
\author{P. M. Kowalski\altaffilmark{1,2}, D. Saumon\altaffilmark{2,1} and S. Mazevet\altaffilmark{3}} 
\affil{\altaffilmark{1} Department of Physics and Astronomy, Vanderbilt University, \\ Nashville, TN 37235, USA}
\affil{\altaffilmark{2} Applied Physics Division, Los Alamos National Laboratory, MS-F699, \\ Los Alamos, NM 87545, USA}
\affil{\altaffilmark{3} Theoretical Division, Los Alamos National Laboratory, MS-B221, \\ Los Alamos, NM 87545, USA}


\begin{abstract} 

The atmospheres of cool, helium-rich white dwarfs constitute an exotic and poorly explored physical regime
of stellar atmospheres. 
Under physical conditions where the temperature varies from $\rm 1000K$ to $\rm 10000K$, 
the density can reach values as large as $\rm 2 \ g/cm^{3}$, and the pressure is as high as $\rm 1 \ Mbar$,
the atmosphere is no longer an ideal gas and must be treated as a dense fluid. 
Helium atoms become strongly correlated and refraction effects are present. 
Opacity sources such as $\rm He^{-}$ free-free absorption 
must be calculated with a formalism that has never been applied to astrophysical opacities.
These effects have been ignored in previous models of
cool white dwarf atmospheres.

\end{abstract}

\section{Introduction}

Very cool white dwarfs (WDs) with $\it T_{\rm eff}\rm<5000K$ are among the oldest objects in our Galaxy. They are of
great importance for cosmochronology and for understanding the formation and evolution of the Milky Way. 
Our main interest is in the atmospheres of these stars as the link between their physical characteristics and the observables. 
In this contribution we concentrate on the atmospheres rich in helium. 
Due to the high transparency of helium (compared to hydrogen), our current models predicts for them 
densities that are typical of liquids, up to $\rm 2-3 \ g/cm^{3}$. 
Under such extreme conditions, the input physics of atmosphere models needs to be reconsidered. 

There are several physical effects that distinguish fluid 
from gaseous atmospheres. They all arise from the strong interaction between particles. Helium atoms become strongly correlated,
the refractive index of the fluid departs from unity, and free electrons strongly interact with the medium, which
affects the opacity.

Non-ideal effects in the equation of state are dominated by $\rm He-He$ interactions. For a given density and temperature, they result 
in increases in the pressure
and ionization fraction and a decrease in the adiabatic gradient. 

The refractive index inside helium-rich white dwarf atmospheres departs from unity and can be as large 
as 1.35. We have solved the equation of radiative transfer for refractive stellar 
atmospheres for the first time (Kowalski \& Saumon 2004). The impact of the total internal reflection on the radiative equilibrium results in 
an increase of the temperature in the optically thin atmospheric layers. This decreases the abundance 
of hydrogen molecules and the $\rm H_{2}-He$ CIA opacity in atmospheres of mixed composition.
Due to strong refraction near the surface, the limb darkening almost disappears. 

Fluid atmospheres require a new description of the opacity that is different from that for diluted gases.
The calculation of the $\rm He^{-}$ free-free absorption (or inverse bremsstrahlung) in WD atmospheres 
must be revised. Preliminary results obtained from quantum molecular dynamics simulations (QMD) suggest that fluid,
helium rich, cool WD atmospheres are much more opaque than predicted by current models.

All these effects must be considered to properly model cool WD atmospheres. 
Realistic models of very cool WDs are of fundamental importance for cosmochronology and to understand the spectra and
physical properties of the coolest WDs, especially the recently discovered, peculiar ultracool stars like LHS 3250.

\section{The Non-Ideal Equation of State}

Our equation of state (EOS) for fluid $\rm H/He$ mixtures includes the following species: 
$\rm H_{2}, H, H^{+}, H^{+}_{2}, H^{+}_{3}, H^{-}, He, He_{2}^{+} ,He^{+}$ and $e^{-}$.
Currently, only the $\rm He-He$ interactions are included as they dominate the non-ideal contributions to the EOS. These interactions
are described with an effective pair potential (Ross \& Young 1986) that is calibrated to high-pressure EOS data \cite{Nellis}.
\begin{figure*}[!ht]
    \plotfiddle{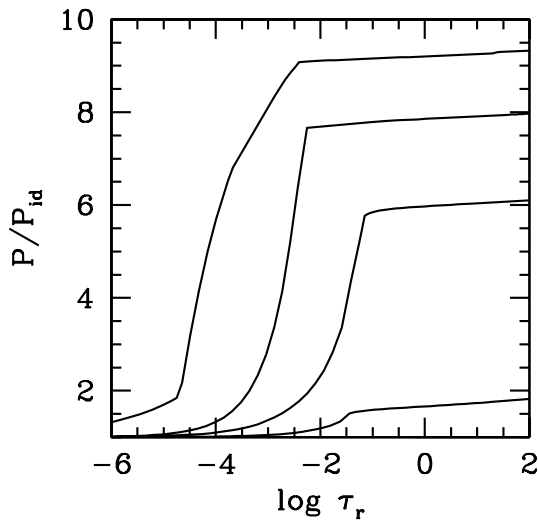}{6.5cm}{0}{60}{60}{-190}{-215}
    \caption {Non-ideal effects in WD atmospheres of $\it T_{\rm eff}\rm=4000K$, 
    $\rm log \ \it g \rm (cm/s^{2}) = 8 $ and various He/H compositions. Starting from the bottom $\rm \it n \rm (He)/\it n\rm(H) = 10^{2}, 10^{4}, 10^{6} \ and \ pure \  He$.
    Shown is the ratio of the total pressure $P$ over the ideal gas pressure $P_{\rm id}$.}
\end{figure*}
The EOS is computed in a manner similar to that of Bergeron, Saumon \& Wesemael (1995) but explicitly includes non-ideal terms in the calculation of the chemical equilibrium
of the H/He mixture. We are therefore able to compute continuous sequences of models from pure H to pure He composition in a self-consistent manner with
the non-ideal EOS. The importance of the non-ideal terms in the EOS is illustrated in Fig. 1. As the He content increases, the overall atmospheric opacity 
decreases and the pressure rises. Clearly, the non-ideal effects dominate the EOS for $\rm \it n \rm (He)/\it n \rm (H) \geq 10^{2}$.  
Because of the repulsive $\rm He-He$ interactions, it is energetically favorable for $\rm He$ to break up into ionized species, and the degree of ionization increases. 

\section{The Refraction}
The variation of the refractive index inside cool WD atmosphere models is displayed in Figure 2.
\begin{figure}[!ht]
    \plotfiddle{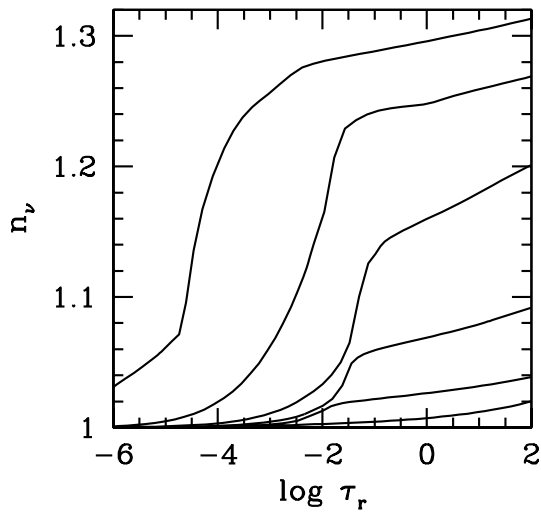}{6.5cm}{0}{60}{60}{-190}{-215}
    \caption {Refractive index inside WD atmospheres of $\rm \it T_{\rm eff}\rm =4000K$, 
    $\rm log \ \it g \rm (cm/s^{2}) = 8 $ and various He/H compositions. Starting from the bottom $\rm \it n \rm (He)/ \it n \rm (H) = 0 \ (pure \ H), 10, 10^{2}, 10^{3}, 10^{5} \ and \ pure \ He$.
    In the pure H atmosphere, the refractive index arises in fluid $\rm H_{2}$.}
\end{figure}

Applying geometric optics we can obtain the radiative transfer equation modified for refraction \cite{Cox}, which has
to be solved along curved, and eventually totally reflected
ray-paths. 
In the atmospheres of cool WDs, the refractive index returns to unity far above the photosphere.
Therefore, the main
effects on refraction are: 1) a significant weakening of limb darkening (from $0.60$ to $0.96$ for pure He models of $\it T_{\rm eff} \rm =4000K$), 2) a reduction of
the opacity, and 3) an increase of temperature in the radiative
zone (Fig. 3). This last phenomenon is a consequence of total internal reflection, which occurs in
the upper atmospheric layers (Kowalski \& Saumon 2004). In mixed H/He models, the larger temperature results in the dissociation of $\rm H_{2}$ 
\begin{figure}[!ht]
    \plotfiddle{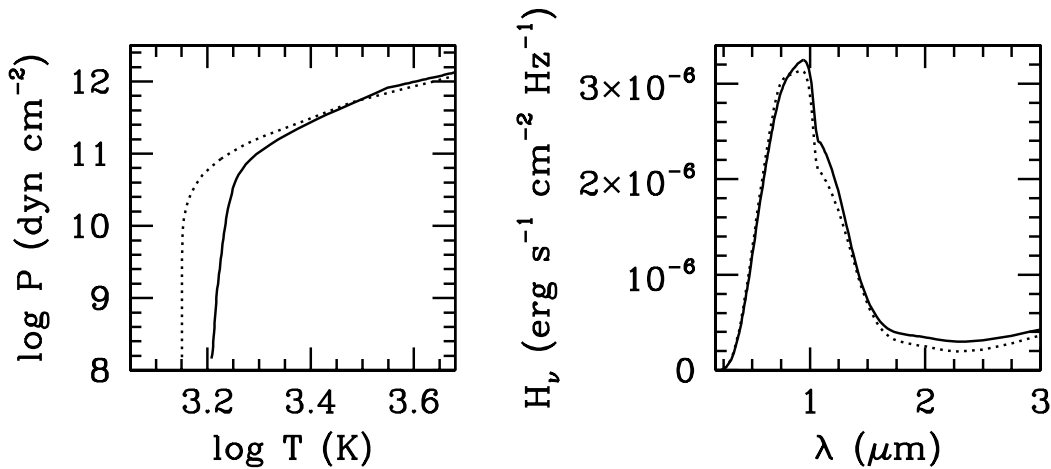}{6.0cm}{0}{70}{70}{-220}{-310}
    \caption {$\it P-T$ structure and synthetic spectrum with (solid line) and without (dotted line) refraction for
    a WD atmosphere model with $\it T_{\rm eff} \rm =4000K$, $\rm log \ \it g \rm (cm/s^{2}) = 8 $ and $\it n \rm (He)/\it n \rm (H)=10^{5}$.}
\end{figure}
and a reduction in the $\rm H_{2}$ CIA absorption in the infrared (Fig. 3). The effect is strongest in the $\it K$ band where the flux is increased by $\rm \sim 30\%$
in this particular model.

\section{The Opacity of Fluid Helium}

The third, and probably the most important density effect in He-rich WD atmospheres is 
its impact on the opacity. We suspect 
that the description of opacities presently used in modeling of He-rich WD atmospheres is inadequate because 
the widely used standard expressions for opacities of H and He species
are valid for tenuous gases, not fluids! The absorption cross section of the important
$\rm He^{-}$ free-free process is derived for an isolated atom-electron collision, while
in a dense fluid, the absorbing electron interacts strongly with many
surrounding helium atoms.

So far, a reduction of
the $\rm He^{-}$ ff absorption and the Rayleigh scattering due to strong, collective interactions 
in dense helium, has been reported (Iglesias, Rogers \& Saumon 2002; hereafter IRS02). This calculation, done in the Born approximation, 
is not quite justified for strong interactions. However, IRS02 show that these two sources of opacity
in dense helium-rich WD atmospheres may be reduced by
factors as large as 20.  

Due to its complexity, a fully analytical, quantum mechanical description of
absorption processes in strongly correlated fluids does not exist. 
However, we can get valuable insight into the opacities of fluid helium
from quantum molecular dynamics simulations \cite{Mazevet}.
The results are surprising. The opacity obtained for pure helium at $\it T \rm=5802K$ 
and $\it \rho \rm = 0.5 \ g/cm^{3}$  is flatter and
much larger than that of IRS02 (Figure 4). Moreover, Kramer's $\rm \nu^{-3}$ behavior of the free-free absorption coefficient is
strongly suppressed. 

\begin{figure}[!ht]
    \plotfiddle{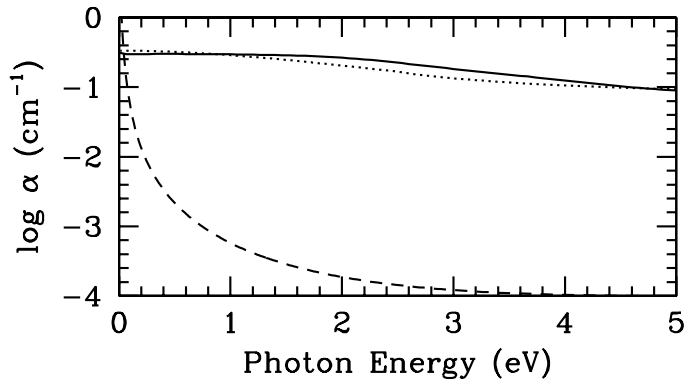}{4.2cm}{0}{60}{60}{-170}{-270}
    \caption {Absorption coefficient $\rm \alpha$ for pure He at $\it T\rm=5802K$ and $\rm \rho =0.5 \ g/cm^{3}$. Lines represent the results from 
     QMD simulations (solid), $\rm He^{-}$ ff of IRS02 and the EOS from \S2 (dashed), 
     and $\rm He^{-}$ ff of IRS02 with $\it f_{\rm e}$ extracted from QMD simulations and the classical model extension for photon energies $\rm \leq \gamma_{c}=2.5 \ eV \sim 12.8 \ \mu m$ (dotted).}
\end{figure}

The frequency behavior of the absorption coefficient obtained from simulations is similar to the prediction of the classical model \cite{Jackson}.
In the classical picture the opacity for small frequencies is asymptotically constant because in the presence of the slowly varying electric field 
of the electromagnetic wave (photon), the mobility of electrons
is determined by the frequency of electron-atoms collisions. 
This effect is present also in quantum systems.
The average frequency of electron-atom collisions $\rm \gamma_{c}$ (damping parameter) for the conditions in Figure 4
is $\rm \gamma_{c}=2.5 \ eV$. 
The ionization fraction $\it f_{\rm e}\rm=\it n\rm (e)/\it n\rm (He)$, extracted from the simulation, by a fit to the classical model
is $\it f_{\rm e}\rm =4.8 \cdot 10^{-7} $, compared to $\it f_{\rm e} \rm =1.8 \cdot 10^{-10} $ 
from our present EOS (\S2). Moreover, extending the $\rm He^{-}$ ff opacity 
with the classical model for frequencies smaller than 
$\rm \gamma_{c}$ and using the ionization fraction extracted from the simulations, we get an opacity that is in
good agreement with the simulations!
This apparently successfull description of the quantum system with a classical model is preliminary and needs to be better understood. Taken at face value,  
these results suggest that multiple electron-atom collisions are important, 
and that the ionization fraction predicted by our EOS is incorrect. 

\section{Implications}
\begin{figure}[!ht]
    \plotfiddle{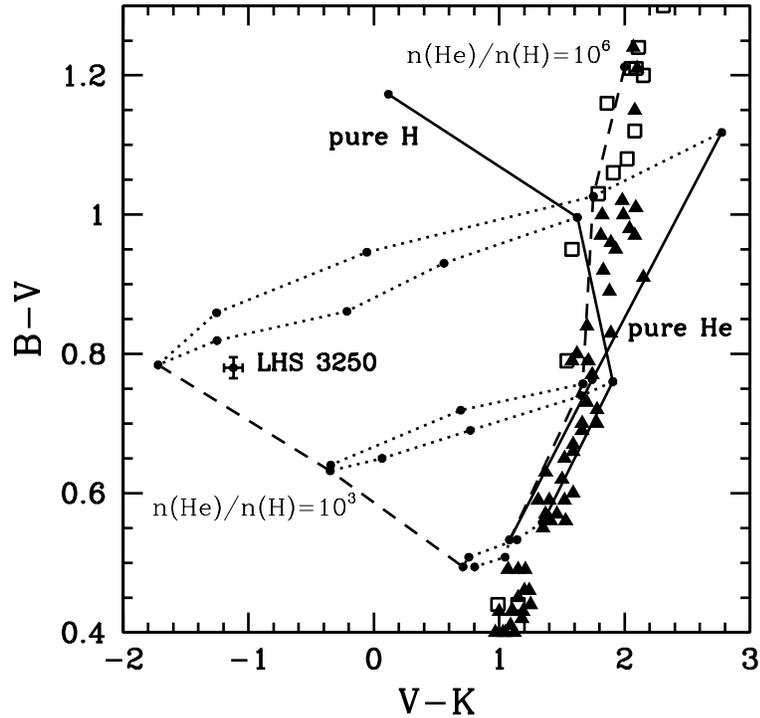}{9.0cm}{0}{75}{75}{-210}{-170}
    \caption {Color-color diagram for cool WDs. Data points come from the BRL97 sample of H-rich (filled triangles) and He-rich (open squares) 
    disk WDs. Sequences of models with mixed $\rm H/He$ composition and constant $\it T_{\rm eff}$ are shown by dotted lines (from top to bottom 
    $\it T_{\rm eff}\rm=4000K,5000K,6000K$). Dashed lines connect models with the same $\rm He/H$ ratio. All models have $\rm log \ \it g \rm (cm/s^{2}) = 8$.}
\end{figure}

Based on the dense fluid physics described here (non-ideal EOS, refraction, and $\rm He^{-}$ ff and Rayleigh scattering of IRS02 but not the QMD opacities), 
we have computed a sequence of 
cool WD atmosphere models (Fig. 5). The comparison with 
the observed sequence of cool WDs (Bergeron, Ruiz \& Leggett 1997; hereafter BRL97)
suggests that the coolest WDs have mixed H/He atmospheres, rather than the pure helium composition assigned to them in BRL97. However,
our prediction for their atmospheric abundance $\it n(\rm He)\rm /\it n(\rm H) \rm \sim10^{6}$
seems too high for disk WDs. Assuming a hydrogen accretion rate of $\rm 10^{-17} M_\odot/year$, 
and a maximal thickness of the convection zone 
of $10^{-3}M_\odot$ \cite{Hansen},
the largest expected abundance of helium is $\it n(\rm He)\rm / \it n(\rm H) \rm \sim10^{4}$. 
Since the QMD simulations predict that pure helium is much more opaque than 
in these models, we anticipate that the fit of the coolest WDs with more complete models will result in a much smaller He/H ratio.

\acknowledgments{Piotr Kowalski is grateful to Vanderbilt University and to the organizers of the 
Workshop 
for financial support which allowed him to attend the meeting. This research 
was supported 
in part by the United States 
Department of Energy under contract W-7405-ENG-36.}

\end{document}